# Precise variational tunneling rates for anharmonic oscillator with $g < 0$ [*]

R. Karrlein and H. Kleinert

Institut für Theoretische Physik
Freie Universität Berlin
Arnimallee 14    D - 1000 Berlin 33

March 9, 1995

## Abstract

We systematically improve the recent variational calculation of the imaginary part of the ground state energy of the quartic anharmonic oscillator. The results are extremely accurate as demonstrated by deriving, from the calculated imaginary part, all perturbation coefficients via a dispersion relation and reproducing the exact values with a relative error of less than $10^{-5}$. A comparison is also made with results of a Schrödinger calculation based on the complex rotation method.

1) It is well known that the evaluation of perturbation expansions can be greatly improved with the help of a variational approach [1]. Recently it was found that also the calculation of tunneling processes can be carried far beyond previously available semiclassical approximation [2] by combining it with variational methods [3]. The range of validity of the semiclassical results was extended far into the regime of low barriers and large tunneling rates. Variational methods allow us to calculate decay rates even when the tunneling barriers are so low that they do not contain any bound states at all. These are the sliding rates of Ref. [3]. For the quantum mechanical anharmonic oscillator with a potential

$$V(x) = \frac{\omega^2}{2}x^2 + \frac{g}{4}x^4 \qquad (\omega^2, g > 0), \qquad (1)$$

low-order calculations of tunneling and sliding rates yield imaginary parts which are so accurate that a dispersion relation produces perturbation coefficients which agree with the exact values calculated by Bender and Wu [4] to *all* orders in perturbation theory.

In this note we want to investigate the increasing accuracy of the approximation method when carrying the systematic improvement scheme of ref.[1] to increasingly higher orders. At the end we shall obtain the imaginary part of the ground state energy with a precision of better than 0.01 %.

[*]Work supported in part by Deutsche Forschungsgemeinschaft under grant no. Kl. 256.

Up to now, the best developed theory of tunneling processes applies only to the limit of high potential barriers or small values of $-g$ where a semiclassical treatment is appropriate. For increasing $-g$, the fluctuation corrections to the semiclassical results become increasingly difficult to calculate. For large $-g$, where the decay proceeds by sliding rather than tunneling, no field theoretic method was available until [3]. The power of the variational method is to obtain accurate results over the entire negative $g$-axis, including the two difficult regimes.

2) The sliding regime was found to be easily calculable. It merely requires an analytic continuation of the variational approximations which is known for positive $g$. There the Rayleigh-Schrödinger perturbation theory yields a power series expansion ($\hbar = 1$)

$$E(g) = \omega \sum_{l=0}^{\infty} e_l^{BW} \left(\frac{g}{4\omega^3}\right)^l \tag{2}$$

where $e_l^{BW}$ are rational numbers

$$\frac{1}{2}, \frac{3}{4}, -\frac{21}{8}, \frac{333}{16}, -\frac{30885}{128}, \ldots \tag{3}$$

The superscript BW refers to Bender and Wu [4] who have found a recursion-relation by which one can easily find the first few hundred terms. The disadvantage of the series (2) is that it has a zero radius of convergence due to the factorial growth of the coefficients $e_l^{BW}$. To be evaluated approximately it has to be truncated at a finite order. The best results are obtained by going to an order $\approx 3/4g$. Reliable numerical values cannot be expected for $g > 0.1$.

The evaluation is greatly improved by a variational treatment. It proceeds as follows: The harmonic term of the potential is split into an arbitrary harmonic term and a remainder:

$$\frac{\omega^2}{2}x^2 \to \frac{\Omega^2}{2}x^2 + \left(\frac{\omega^2}{2} - \frac{\Omega^2}{2}\right)x^2 \tag{4}$$

Writing

$$V(x) = \frac{\Omega^2}{2}x^2 + V_{\text{int}}(x) \tag{5}$$

with an interaction

$$V_{\text{int}}(x) = \frac{g}{4}(rx^2 + x^4), \quad r = \frac{2}{g}(\omega^2 - \Omega^2) \tag{6}$$

one performs perturbation expansion in powers of $g$ at a fixed $r$:

$$E_k(g,r) = \Omega \sum_{l=0}^{k} e_l(r) \left(\frac{g}{4\Omega^3}\right)^l \tag{7}$$

The calculation of the new series up to a specific order $k$ does not require much additional work since it is easily obtained from the ordinary perturbation series (2) by replacing $\omega$ by $\sqrt{\Omega^2 + gr/2}$ and expanding in powers of $g$ up to the $k$th order. This yields

$$e_l(r) = \sum_{j=0}^{l} e_j^{\text{BW}} \binom{(1-3j)/2}{l-j} (2r\Omega)^{l-j} \tag{8}$$

The truncated power series

$$W_k(g, \Omega) := E_k\left(g, 2(\omega^2 - \Omega^2)/g\right) \tag{9}$$

is certainly independent of $\Omega$ in the limit of large $k$. At any finite order $k$, however, it does depend on $\Omega$, the approximation having its fastest speed of convergence where it depends least on $\Omega$. Denote the order-dependent optimal value of $\Omega$ by $\Omega_k(g)$. The quantity $W_k(g, \Omega_k(g))$ is the new approximation to $E(g)$ [5]. Since $W_k(g, \Omega)$ is of the form $\sum_{m=0}^{3k} a_m(g)\Omega^{m-3k+1}$ with polynomials $a_m(g)$, the defining equation for its extrema is a polynomial of order $3k$ in $\Omega$. Thus only in the case of odd $k$, a minimum is guaranteed to exist. In the case of even $k$, the alternating sign of $a_0(g)$ turns out to remove the minimum for smaller $k$. Not considering very small positive $g$ (for which there always exist $k$ real solutions in the vicinity of $\omega$) it has been found numerically that $W_k(g, \Omega)$ has no extrema for $k = 2, 4$, one extremum for $k = 1, 3, 5, 7, 9, 11, 13$, two extrema for $k = 6, 8, 10, 12, 14, 16, 18, 20$ and three extrema for $k = 15, 17, 19, 21$. At further increasing orders the situation becomes increasingly difficult to handle numerically because the numerous extrema are to close. An interesting situation occurs in the case of three extrema when there are two minima. The global minimum of $W_k(g, \Omega)$ turns out to yield a 100-times less accurate energy value than the higher lying local minimum. In Table 1 we display the values obtained from the better local minimum in the cases $k = 15$ and $k = 21$. In this way, an agreement of 0.01 ppm with the values given in the literature [6] was reached. An interesting fact is that the inequality $E_{\text{exact}}(g) \leq \text{Min}\,(W_k(g,\Omega))$ which in the case $k = 1$ follows from the Jensen-Peierls inequality seems to hold also for $k = 3$, whereas it is definitely violated for higher $k$. Note, however that the above-discussed preferable local minima do again provide upper bounds in accordance with the Jensen-Peierls inequality.

Our approximation is easily generalized to the excited states. One only has to replace the ground state perturbation coefficients by the corresponding coefficients $e_{l,n}^{\text{BW}}$ for the $n$th excited state. We have calculated them up to $k = 5$ for all $n$ using usual Rayleigh-Schrödinger perturbation theory. The error of the resulting approximation behaves similar to that of the ground state energies.

| $g/4$ | $E_{\text{exact}}(g)$ | $k=1$ | $k=3$ | $k=5$ | $k=7$ | $k=11$ | $k=15$ | $k=21$ |
|---|---|---|---|---|---|---|---|---|
| 0.1 | 0.5591463 | 0.5603074 | 0.5591542 | 0.5591462 | 0.5591462 | 0.5591462 | 0.5591463 | 0.5591463 |
| 0.3 | 0.6379918 | 0.6416299 | 0.6380358 | 0.6379899 | 0.6379897 | 0.6379903 | 0.6379918 | 0.6379918 |
| 0.5 | 0.6961758 | 0.7016616 | 0.6962536 | 0.6961717 | 0.6961712 | 0.6961723 | 0.6961758 | 0.6961758 |
| 1.0 | 0.8037706 | 0.8125000 | 0.8039140 | 0.8037615 | 0.8037601 | 0.8037623 | 0.8037707 | 0.8037706 |
| 2.0 | 0.9515685 | 0.9644036 | 0.9517998 | 0.9515517 | 0.9515490 | 0.9515526 | 0.9515685 | 0.9515685 |
| 50 | 2.4997088 | 2.5475804 | 2.5006996 | 2.4996213 | 2.4996061 | 2.4996219 | 2.4997089 | 2.4997088 |
| 200 | 3.9309313 | 4.0084609 | 3.9325538 | 3.9307858 | 3.9307603 | 3.9307863 | 3.9309316 | 3.9309313 |
| 1000 | 6.6942208 | 6.8279533 | 6.6970329 | 6.6939668 | 6.6939221 | 6.6939673 | 6.6942213 | 6.6942209 |
| 8000 | 13.366908 | 13.635283 | 13.372561 | 13.366395 | 13.366305 | 13.366396 | 13.366908 | 13.366908 |
| 20000 | 18.137229 | 18.501659 | 18.144908 | 18.136533 | 18.136411 | 18.136534 | 18.137230 | 18.137229 |

Table 1: Comparison of the exact values of $E(g)$ with those obtained by minimization of $W_k(g,\Omega)$.

To judge the quality of our approximation consider the asymptotic strong-coupling behavior of $E(g)$ both for negative as for positive $g$ which's exact numerical values are known with great precision [6]. It can be parametrized as follows:

$$E_n(g) = (g/4)^{1/3}\left[\alpha_n + \beta_n(g/4)^{-2/3} + \gamma_n(g/4)^{-4/3}\ldots\right]. \qquad (10)$$

The asymptotic behavior of $\text{Im}E_n(-g)$ is obtained by an analytic continuation:

$$\text{Im}E_n(-g) = (g/4)^{1/3}\left[\alpha'_n + \beta'_n(g/4)^{-2/3} + \gamma'_n(g/4)^{-4/3}\ldots\right], \qquad (11)$$

| $k$ | $\alpha_0$ | $\beta_0$ | $k$ | $\alpha_0$ | $\beta_0$ |
|---|---|---|---|---|---|
| 1 | 0.681420222 | 0.13758 | 9 | 0.667958214 | 0.143700 |
| 3 | 0.668269437 | 0.14347 | 11 | 0.667960613 | 0.143698 |
| 5 | 0.667960581 | 0.14369 | 13 | 0.667962642 | 0.143696 |
| 7 | 0.667956063 | 0.14370 | 15 | 0.667986302 | 0.143669 |
|   |   |   | exact | 0.667986259 | 0.14367 |

Table 2: Comparison of the exact asymptotic behavior with those obtained by our approximation at increasing order $k$ of the variational approximation.

with coefficients

$$\alpha'_n = \frac{\sqrt{3}}{2}\alpha_n \; , \quad \beta'_n = -\frac{\sqrt{3}}{2}\beta_n \; , \quad \gamma'_n = -\gamma_n \; . \tag{12}$$

Table 2 shows the values of our approximation for the ground state coefficients $\alpha_0$ and $\beta_0$ in comparison with the exact values. The agreement is seen to improve rapidly with the order $k$ of the approximation. Our approximation for the coefficients $\alpha_n$ and $\beta_n$ of the excited states is displayed in Table 3. The order of the approximation is $k = 5$.

3) It was shown in ref. [3] for the lowest case $k = 1$, that the above approximation scheme makes also sense for sufficiently negative $g$. There the imaginary part of the ground state energy is obtained by an analytic continuation of the minimum of $W_1(g,\Omega) = 3g/16\Omega^2 + 1/4\Omega + \Omega/4$. For $g < -4/3^{5/2} \approx -0.25$ the optimal value of $\Omega$ moves into the complex $g$-plane and the imaginary part of the energy describes a sliding process into the abysses. In the case of higher (odd) orders $k$, this analytic continuation has to be done numerically. There are standard computer programs which give all the $3k$ roots of $dW_k(g,\Omega)/d\Omega$ for arbitrary complex $g$ to any precision. From these we have to choose the one that is continuously connected (i.e. on a semicircle in the complex g-plane) with the best (in the sense as discussed above) solution on the positive $g$-axis. With increasing orders the range of validity of the approximation moves closer and closer to the origin. This is quite amazing since the behavior of $\text{Im}E(g)$ in the limit of small $-g$ possesses an essential

| $n$ | $\alpha_n$ | $\beta_n$ |
|---|---|---|
| 0 | 0.6679605808 | 0.1436962401 |
| 1 | 2.3922963623 | 0.3586118166 |
| 2 | 4.6688449582 | 0.5059561947 |
| 3 | 7.2862672627 | 0.6340450238 |
| 4 | 10.171684571 | 0.7501258899 |
| 5 | 13.282179566 | 0.8577545339 |
| 6 | 16.588882965 | 0.9589691819 |
| 7 | 20.070767345 | 1.0550754253 |
| 8 | 23.711695355 | 1.1469734787 |
| 9 | 27.498813862 | 1.2353174155 |
| 10 | 31.421596541 | 1.3206014789 |

Table 3: Coefficients governing the asymptotic behavior of the excited states at a fixed order $k = 5$ of the variational approximation.

singularity

$$\mathrm{Im} E^\omega_{\mathrm{sc}}(g) = \omega \sqrt{\frac{6}{\pi}} \sqrt{\frac{4\omega^3}{3|g|}} e^{-4\omega^3/3|g|}. \tag{13}$$

Our analytic approximants approach rapidly this nonanalytic behavior. Figure 1 shows the approximations at different orders. We have plotted the reduced imaginary part $\varepsilon(g) \equiv \mathrm{Im} E(g)/\mathrm{Im} E_{\mathrm{sc}}(g)$ whose semiclassical limit is unity.

To check the correctness of our result we compare it with the values given in the literature which have an accuracy of 0.1 % [7]. We derive even more precise values by using the so-called "complex rotation method" [10]. It yields a matrix recurrence relation which can be solved using matrix continued fractions (see ref. [11]). This method yields numerical values which coincide with those of our approximation within a relative error of 0.1%. Table 4 compares the three independently obtained results. Our values have about the same precision as those obtained by Drummond. This can be seen from the results of the "complex rotation method" which are calculated to very high accuracy.

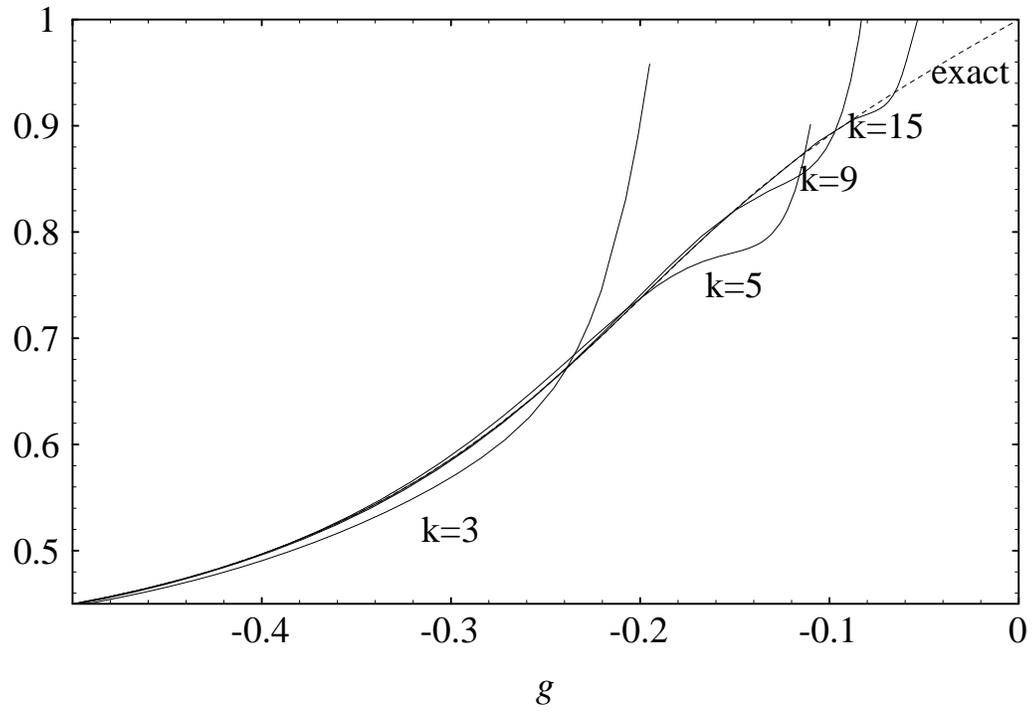

Figure 1: Reduced imaginary part at different orders of our approximation

| $g$ | $\varepsilon_{\text{sc}}^{\text{var}}$ | $\varepsilon_{\text{sc}}^{\text{D}}$ | $\varepsilon_{\text{sc}}^{\text{coro}}$ |
|---|---|---|---|
| -0.08 | 0.9108491971 | 0.92 | 0.9149958646 |
| -0.1 | 0.8929234815 | 0.8907 | 0.8910391177 |
| -0.12 | 0.8650453831 | 0.8650 | 0.8650720065 |
| -0.14 | 0.8356618164 | 0.8357 | 0.8364609072 |
| -0.16 | 0.8049117571 | 0.8051 | 0.8050709206 |
| -0.18 | 0.7718751164 | 0.7714 | 0.7715176248 |
| -0.2 | 0.7373677436 | 0.7366 | 0.7369712418 |
| -0.24 | 0.6699084973 | 0.6698 | 0.6699503639 |
| -0.3 | 0.5859071721 | 0.5862 | 0.586125402 |
| -0.4 | 0.4967562191 | 0.4966 | 0.4966786117 |
| -0.5 | 0.4498031025 | 0.4496 | 0.4496651302 |
| -0.6 | 0.4258522448 | 0.4258 | 0.4257819772 |
| -0.8 | 0.4114658085 | 0.4115 | 0.4115284431 |
| -1.0 | 0.4177169144 | 0.4178 | 0.4178285991 |
| -1.2 | 0.4332072064 | 0.4333 | 0.4333195993 |
| -1.4 | 0.4534503231 | 0.4535 | 0.4535431192 |
| -1.6 | 0.4763745530 | 0.4764 | 0.4764405650 |
| -1.8 | 0.5009071563 | 0.5009 | 0.5009448702 |
| -2.0 | 0.5264438875 | 0.5265 | 0.5264541474 |

Table 4: Comparison of our approximation of the reduced imaginary part $\varepsilon(g)$ (order $k = 15$) with Drummond's results and the precise numerical values obtained via the solution of the Schrödinger equation with the complex rotation method.

4) In the vicinity of the origin where the decay proceeds by tunneling, the previous approximation gives no imaginary part. But we can adjust it to the semiclassical behavior as follows: One first has to calculate corrections to the semiclassical behavior

$$\mathrm{Im} E(g) = \mathrm{Im} E_{\mathrm{sc}}^{\omega}(g) \exp\left[-\sum_{j=1}^{k} d_j (-\frac{3|g|}{4\omega^3})^j\right] \qquad (14)$$

where $d_j$ are positive rational numbers. The first of them can be calculated using field theoretic techniques [9]. In the special case of the quartic anharmonicity Zinn-Justin [8] has found them numerically up to $d_{10}$. Unfortunately, the power series in $g$ is only asymptotic one so its accuracy does not increase if $k$ is carried beyond $\approx 4/3|g|$. The variational approach improves the results and the convergence dramatically since the effective coupling constant is the dimensionless quantity $\alpha(g) = g/\Omega^3(g)$ which is always smaller than 0.6 this value being reached only in the strong-coupling limit $g \to \infty$ (this limit holds for $k = 1$; for $k = 3$ the maximal value decreases to 0.3. For a plot see [12]).

To improve this variationally one rewrites this by introducing a trial frequency $\Omega$ as

$$\mathrm{Im} E(g) = \mathrm{Im} E_{\mathrm{sc}}^{\Omega}(g) \exp\left[\frac{5}{2}\log(\frac{\omega}{\Omega}) - \sum_{j=-1}^{k} d_j (-\frac{3|g|}{4\omega^3})^j\right] \qquad (15)$$

with

$$d_{-1} = 1, \quad d_0 = 0, \quad d_1 = \frac{95}{72}, \quad \ldots \qquad (16)$$

At this point we replace $\omega$ by $\sqrt{\Omega^2 + g/2r}$,   ( $r = 2(\omega^2 - \Omega^2)/g$ ) and expand in powers of $g$ at fixed $r$ up to order $k$ which yields

$$\mathrm{Im} E(g) = \mathrm{Im} E_{\mathrm{sc}}^{\Omega}(g) \exp\left[\frac{5}{4}\sum_{j=1}^{k} \frac{1}{j}(\frac{\omega^2 - \Omega^2}{\Omega^2})^j - \sum_{j=-1}^{k} d_j(\Omega)(-\frac{3|g|}{4\omega^3})^j\right] \qquad (17)$$

where

$$d_j(\Omega) = d_j \sum_{m=0}^{k-j} \binom{-3j/2}{m} (\frac{\omega^2 - \Omega^2}{\Omega^2})^m (\frac{\omega}{\Omega})^{3j}. \qquad (18)$$

$\Omega$ is still determined by an analytic continuation of the above optimal $\Omega_1$ for all $k$. For $k = 1$ we obtain therefore

$$\text{Im}E(g) = \text{Im}E_{\text{sc}}^{\Omega}(g) \exp\left[\frac{5}{4}x + (1 + \frac{3}{2}x + \frac{3}{8}x^2)\frac{4\Omega^3}{3g} + \frac{95}{72}\frac{3g}{4\Omega^3}\right] \qquad (19)$$

where we have used the abbreviation $x = (\omega^2 - \Omega^2)/\Omega^2$. The resulting approximation is shown in Figure 2 for $k = 1$ and $k = 2$.

The same reasoning can also be used for the excited states. Difficulties arise only in the calculation of $d_{k,n}$ for arbitrary $n$ since the field theoretic method of [9] cannot be applied. There exist however a WKB calculation of $d_{1,n}$ [4]. Figure 3 shows these variational approximations for the ground state and the first two excited states. We have also plotted the usual approximation for $k = 1$ and $k = 3$ to show how well the two approximations fit together. Figure 1 and 2 are magnifications of the dashed box.

We can now use our knowledge of the entire cut to calculate the perturbation coefficients $e_l^{\text{BW}}$ by inserting it into the dispersion relation

$$e_l^{\text{BW}} = \frac{1}{\pi}\int_{-\infty}^{0} \text{Im}E(g)g^{-l-1}dg. \qquad (20)$$

If we take for example $k = 15$ we can use our approximation left of $-2/15 \approx 0.133$ and Eq. (14) truncated after $k = 10$ to continue it up to $g = 0$. The results are given in Table 5 together with the values for $k = 3$ (which were calculated in combination with the semiclassical result as in ref.[3]). They show again the quality of our approximation.

5) The calculations presented in this paper were only intended as a preparation for finding the imaginary part of the correlation functions of quantum field theories. If we succeed to extend our method to the $\varphi^4$-theory we would be able to study its properties with a much higher accuracy than presently available on the basis of standard resummation methods. In particular, there is hope that the precision of renormalization group calculations of critical indices can greatly be improved.

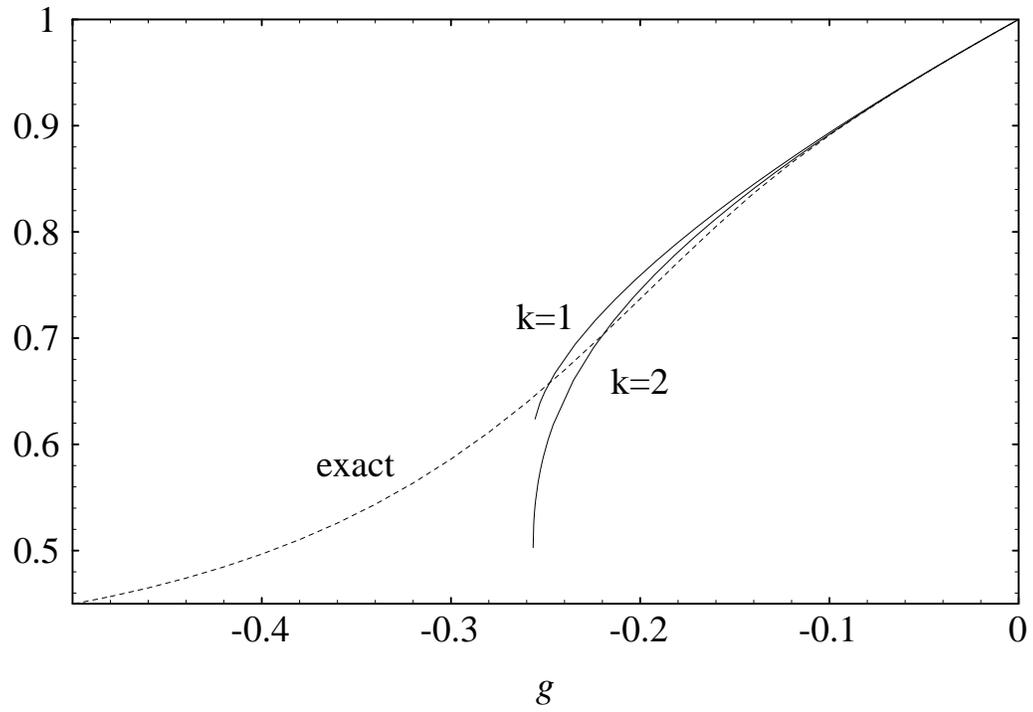

Figure 2: Approximation for the reduced imaginary part in the tunneling regime

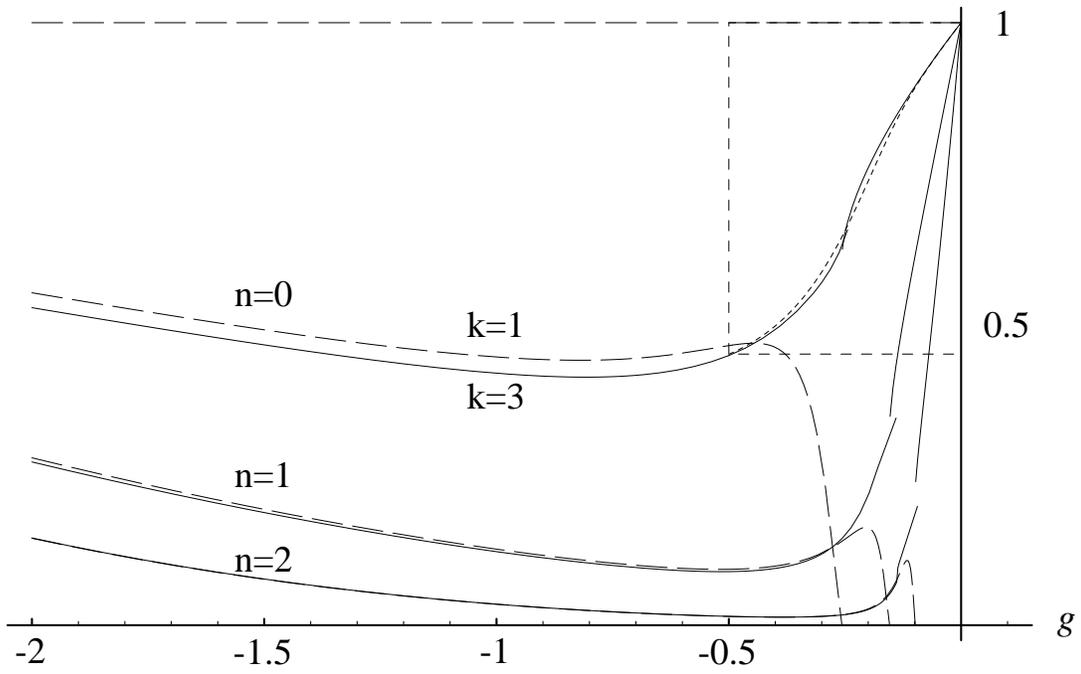

Figure 3: Approximation of the reduced imaginary part of the ground state and the first two excited states

| $l$ | $E_l^{\text{BW}}$ | $k=3$ | $k=15$ |
|---|---|---|---|
| 1 | 0.75 | 0.749321679 | 0.750035087 |
| 2 | 2.625 | 2.614620115 | 2.624986293 |
| 3 | 20.8125 | 20.7186128 | 20.81251773 |
| 4 | 241.289063 | 240.857317 | 241.289495 |
| 5 | 3580.98047 | 3590.69587 | 3580.97722 |
| 6 | 63982.8135 | 64432.53875 | 63982.3799 |
| 7 | 1329733.73 | 1342857.03 | 1329709.29 |
| 8 | 31448214.7 | 31791078.07 | 31447170.8 |
| 9 | 833541603 | 842273537 | 833504391 |
| 10 | 24478940700 | 24703889150 | 24477781497 |

Table 5: The exact perturbation coefficients and those calculated from our approximation and the dispersion relation.

# References

[1] H. Kleinert, Phys. Lett. A *173*, 332 (1993)

[2] A. I. Vainshtein, *Decaying Systems and the Divergence of Perturbation Series*, Novosibirsk Report (1964), im Russian (unpublished)

   J. S. Langer, Ann. Phys. *41*, 108 (1967)

[3] H. Kleinert, Phys. Lett. B *300*, 261 (1993)

[4] C.M. Bender and T.T. Wu, Phys. Rev. *184*, 1231 (1969), Phys. Rev. D *7*, 1620 (1973)

[5] Note that our approximation is far more accurate than the one used in many papers on the acceleration of convergence of perturbation theory such as
Yamazaki, J. Phys. A *17*, 345 (1984)
W.E. Caswell, Ann. Phys. *123*, 153 (1979)
The approximations in these papers are reproduced if we use for all $k$ $\Omega = \Omega_1$ as a variational frequency.

[6] F.T Hioe and E. W. Montroll, J. Math. Phys. *16*,1945 (1975)
    The most accurate energies have been calculated by
    F. Vinette and J. Čižek, J. Math. Phys. *32*, 3392 (1991)

[7] J.E.Drummond, J.Phys.A : Math. Gen. *14* 1651 (1981), J.Phys.A : Math. Gen. *15* 2321 (1982) He calculates the imaginari parts by summing the asymptotic series by means of a truncated binomial model. In the second paper he compares them with a Schrödinger calculation using a Runge-Kutta algorithm.

[8] J. Zinn-Justin, J. Math. Phys. *22*, 511 (1981)

[9] J.C. Collins and D.E. Soper, Ann. Phys. *112*, 209 (1978)

[10] N. Moiseyev and J. O. Hirschfelder, J. Chem. Phys. *88*, 1959 (1988) and references therein

[11] H. Risken, *The Fokker-Planck Equation*, Berlin, Heidelberg: Springer (1984), Springer Series in Synergetics Vol.18 (chapter 9)

[12] H. Kleinert and H. Meyer, Berlin Preprint Oct. 93